# Abnormal Grain Growth of Rutile TiO$_2$ Induced by ZrSiO$_4$

**Dorian A.H. Hanaor**[*]**, Wanqiang Xu, Michael Ferry and Charles C. Sorrell**

*School of Materials Science and Engineering, University of New South Wales, Sydney, NSW 2052, Australia*

*Corresponding Author: Email dorian@unsw.edu.au     Ph: (+61)404188810

---

## Abstract

Abnormal grain growth was observed in rutile TiO$_2$ formed by the thermal treatment of anatase TiO$_2$ in the presence of zirconium silicate. This morphological behaviour was seen to occur in sintered powder compacts and thin films with solid state zircon dopants and in TiO$_2$ coatings on grains of zircon sand. In order to clarify the mechanism by which this grain growth is caused, various doping methods were employed and the morphological consequences of these doping methods were investigated. It was found that doping by Zr and/or Si does not give rise to abnormal grain growth. The observed phenomena were discussed in terms of morphological and energetic considerations.  It is likely that a distinct orientation relationship between rutile TiO$_2$ and ZrSiO$_4$ and possible grain boundary liquid formation play a role in giving rise to the rapid growth of faceted prismatic rutile.

---

# 1.  Introduction

The phenomenon of abnormal grain growth (AGG), also known as exaggerated grain growth, involves the rapid growth of certain grains in a matrix of finer grains exhibiting much slower growth rates [1, 2]. This phenomenon occurs in various systems including ($Al_2O_3$) [3-6] and barium titanate (0.1 mol% Ti excess $BaTiO_3$) [7-9].  AGG in single and multi-phase systems is generally considered to require a high level of anisotropy in surface energy, thus favouring rapid growth in certain crystallographic directions. A high level of chemical inequilibrium, an initially bimodal distribution of grain sizes and high local rates of atomic migration are also reported to play an important role in this phenomenon[1, 2].

A theory of abnormal grain growth was proposed by Hillert [10]. Although focusing principally on multi phase metallic systems, it was suggested that the presence of insoluble pinning particles within a certain size range is responsible for abnormal grain growth in multi phase systems. In ceramic systems various other mechanisms through which abnormal grain growth may arise have been proposed.

Single phase $BaTiO_3$ has been observed to exhibit abnormal growth of grains containing {111} lamellae in certain crystallographic orientations through the lateral movement of faceted grain boundaries [8]. Faceted grain boundaries were shown to be a requisite condition for the occurrence of AGG in this system, an observation which is consistent with other reports which suggest that grain boundary faceting is necessary for AGG in ceramic materials [11]. Indeed defaceting of grain boundaries can be used to prevent the occurrence of AGG [4]. Apart from barium titanate, AGG in ceramic systems is reported only for systems consisting of two or more phases (with dopant or liquid phases present).

In ceramic systems containing two or more phases, the presence of a liquid phase at grain boundaries due to incipient melting or fluxing impurities has been reported to result in the abnormal grain growth of faceted grains. This has been reported in the case of doped alumina materials[3], Barium Titanate [12] and Strontium Barium Niobate [13]. AGG is enhanced as a result of grain boundary liquid phase due to the more rapid diffusion kinetics and high grain boundary mobility.

In a study by Recnik et al., ceramic systems were found to exhibit AGG without the formation of liquid due to polytype interfaces between solid phases [14]. This study reported that polytypic faults in sintered $CaTiO_3$ perovskite doped with BaO caused abnormal grain growth in the direction of fault planes.

Materials which exhibit abnormal grain growth may exhibit divergent properties to materials of a similar composition with homogenous grain sizes.  Often abnormal grain growth is considered to be undesirable due to adverse effects that such growth may have on hardness of the material through Hall-Petch behaviour [15]. Several papers have reported steps that can be taken to prevent the occurrence of AGG in ceramic materials in order to achieve full density and improved mechanical properties [16, 17].

In contrast, abnormal grain growth may, in certain circumstances, have a beneficial effect on mechanical properties. It was found that subsequent to liquid phase sintering, silicon carbide which had undergone abnormal grain growth exhibited improved fracture toughness as a result of crack-tip/crack-wake bridging by elongated abnormally large grains [18-22]. As silicon carbide is used in ceramic armour, enhanced fracture toughness is of great importance in this material [23]. Similar AGG-toughening effects were observed in doped β-phase Silicon Nitride ($Si_3N_4$) [24-26] and have been proposed as a method for toughening alumina [27]. This type of crack-bridging based enhanced fracture toughness of ceramic materials exhibiting AGG is consistent with reported morphological effects on crack propagation in ceramics [28, 29].

Titanium dioxide, which is primarily used as a pigment, has been attracting increasing attention for its potential in photocatalysis, utilising solar irradiation to facilitate environmental remediation through the destruction of contaminants in water. The anatase-rutile phase assemblage and grain morphology in $TiO_2$ has significant consequences on the photocatalytic performance of this material [30-32], and for this reason $TiO_2$ exhibiting AGG of the rutile phase is of interest. To date AGG has not been reported in $TiO_2$. Further potential benefits of AGG in $TiO_2$ may include improved fracture toughness in sintered material, potentially broadening the scope of $TiO_2$ applications.

The current work reports the abnormal grain growth of rutile $TiO_2$ which takes place when anatase $TiO_2$ is treated at high temperatures in the presence of Zircon, $ZrSiO_4$. The phenomena reported in the present work present a new approach to controlling the morphology of mixed and single phase $TiO_2$.

## 2. Experimental Procedures

### 2.1. Powder compacts

Undoped powder compacts were prepared from anatase powder (>99.5%, Merck Chemicals, Australia) ground in an agate mortar and pestle and subsequently pressed in a uniaxial press at 30 kN (95.5 MPa) to form 20 mm diameter powder compacts. Zircon doped samples were made following the same procedure with the addition of 1-15 wt% zircon flour (300 Mesh, Wallarah Minerals, Doyalson, NSW). These powder compacts were fired in air for 4h at 1025°C with a heating rate of 2°/min. Firing durations were extended to 12h to examine microstructure development.

For comparison purposes powder compacts were prepared from anatase powder blended with 10.6 wt% monoclinic zirconia $ZrO_2$ (Z-Tech, USA) to give the same Zr content as the 15 wt% zircon flour doped sample (7.14 at%). This was done to determine whether the observed growth habit was a result of Zr diffusion into the $TiO_2$ lattice. $ZrO_2$ doped samples were fired for 2h at 1025°C as it was found that this resulted in an anatase:rutile ratio similar to that of the zircon doped material.

To further examine Zr and Si doping on the grain growth morphology of $TiO_2$, powders were prepared with dopants introduced using a wet impregnation method. In this method $TiO_2$ powders were suspended in 0.2 M suspensions of zirconium isopropoxide and tetra-ethyl-orthosilicate(TEOS) (both Sigma Aldrich, USA) in isopropanol, and subsequently dried in air. Using this method $TiO_2$ powders doped with 10 at% Si, 10 at% Zr and co-doped with 5 at% Si/Zr were prepared. These dopant-impregnated powders were fired for 4h at 1025°C in an electric muffle furnace with a heating rate of 2°/minute.

### 2.2.    Thin films

Undoped and zircon flour doped $TiO_2$ thin films on quartz substrates were prepared by spin coating using a sol-gel precursor, in a method similar to that reported elsewhere [33]. A solution of 0.5M titanium tetra-isopropoxide (TTIP) (Sigma Aldrich, USA) in isopropanol (100%) was hydrolysed under stirring with the dropwise addition of 2M HCl giving a molar ratio of 2:1 $H_2O$:Ti. Thin films containing zircon flour were prepared using a 0.03g/ml suspension of zircon flour in the sol-gel described above. Films were cast on 20mm x 20mm x 1mm quartz substrates using a Laurell WS-650Sz spin coater at 1000RPM. Films were fired in a muffle furnace at 900 °C for 12 hours with a heating rate of 2°/minute in order to achieve complete transformation to the rutile phase. Owing to the use of a single deposition, deposited $TiO_2$ layers were too thin to form fully dense films, which are easily obtained by the spin coating of more viscous sol or by the use of multiple depositions [33].

### 2.3.    Coated Sand

High surface area supported photocatalysts were prepared through the application of sol-gel coatings of $TiO_2$ on zircon sand and quartz sand (both Wallarah minerals, Doyalson, NSW). These materials were made by suspending 50g of sand in a 50ml rapidly stirring 0.5M TTIP solution in isopropanol. Distilled water was added to give a 6:1 hydrolysis ratio which resulted in the precipitation of colloidal amorphous $TiO_2$ particles. This was followed by the sedimentation of the sand in the suspension and the removal of excess precipitates from the supernatant subsequent to solvent evaporation.  Coated sands were fired in air at 1000 °C for 4 hours with a heating rate of 2°/minute to give rutile $TiO_2$ coatings on the sand grains.

### 2.4.    Coatings on facets of single crystals of zircon

To determine the effects of different $ZrSiO_4$ crystallographic planes on the grain growth of rutile, $TiO_2$ coatings were applied to polished surfaces of a sectioned single crystal of zircon (Peixes, Brazil). Single crystal X-ray Diffraction (XRD) was carried out using a Bruker Kappa Apex Single Crystal Diffractometer to determine the crystallographic planes present in the sections. A single crystal colourless $ZrSiO_4$ gemstone (Aurora, USA) was further used without sectioning. Coatings were applied to crystal surfaces by the dropwise addition of 0.5M TTIP solution in isopropanol and allowed to dry in ambient air. Coated single crystal surfaces were then fired at 1000°C for 4h in air with a heating rate of 2°/minute.

*2.5.    Analysis*

The contaminants present in raw materials used were analysed using X-Ray Fluorescence (XRF) using a Philips PW2400 XRF spectrometer.

Phase analysis of powders was carried out by X-ray diffraction (XRD)using a Phillips multi-purpose diffractometer (MPD) system. Phase content was calculated using the method of Spurr and Myers [34].

Microstructural analysis was performed by Scanning electron microscopy (SEM) using Hitachi S3400, S4500 and S900 microscopes and high resolution SEM (HRSEM) using a FEI Nova-NanoSEM 230. Samples were sputter-coated with chromium to facilitate conductivity. Uncoated specimens of coated single crystal facets were analysed using a Hitachi TM3000 table-top SEM.

Phase analysis with reference to microstructural features in sintered powder compacts was facilitated by laser Raman microspectroscopy in conjunction with optical microscopy using an Invia Raman unit with excitation by a 514 nm wavelength laser. This system allows the interpretation of phase assemblage with reference to microscopically visible features. Laser Raman microspectroscopy was used also for the mineralogical analysis of $TiO_2$ coatings on sand as XRD analysis was inappropriate for such materials due to the dominance of peaks resulting from the sand support material.

# 3.  Experimental Results

*3.1.    Powder compacts*

XRD patterns gathered from crushed sintered powder compacts were interpreted using the method of Spurr and Myers [34] to determine the fraction of the retained anatase phase. This is shown in **Fig. 1a.** Undoped anatase shows complete transformation to rutile after 4h of firing at 1025°C. Under the same conditions, zircon doped samples show significant anatase retention, as seen also in the diffractograms shown in **Fig. 1b**. The inhibition of the anatase to rutile phase transformation in the presence of zircon a dopant is likely the result of the diffusion of Si and Zr in the $TiO_2$ lattice, or grain boundaries, and the consequent restriction of the atomic rearrangement and grain growth involved in the phase transformation as discussed elsewhere [35-38].

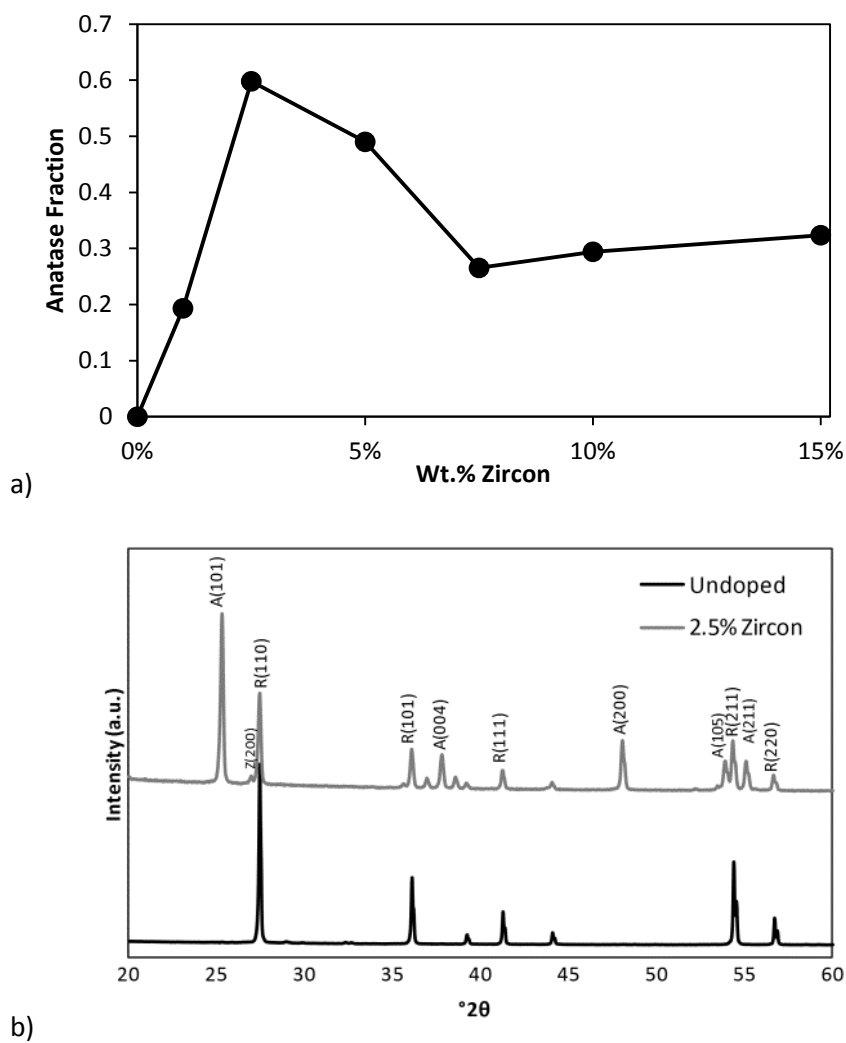

a)

b)

Fig. 1. (a) Retained anatase fraction in powder compacts doped with zircon flour fired at 1025°C for 4h. (b) X-ray diffractograms from undoped and 2.5 wt% $ZrSiO_4$ doped samples with anatase rutile and zircon peaks marked.

Sintered undoped $TiO_2$ samples exhibit rounded isotropic grains of rutile, exhibiting sizes in the region 200-500 nm. This is shown in **Fig. 2a.** This morphology of rutile formed from the thermal treatment of anatase in the absence of dopants is consistent with reported results from the thermal treatment of similar precursor material [39, 40]. In contrast, SEM analyses of zircon flour doped sintered powder compacts showed large elongated prismatic grains growing in close proximity to zircon particles as shown in **Fig. 2b.** These grains exhibited significantly larger sizes approximately 1-5 μm in width and 5-30 μm in length and the morphology was resemblant of AGG in other ceramic materials shown in the literature [21]. As illustrated by **Fig. 3**, these large elongated grains formed around zircon particles and subsequent to extended firing (12 h) such grains had grown throughout the entire sample. Growth of elongated prismatic grains progressed until the point when such grains impinged on each other. As the anatase particles were not present in a preferred orientation, the formation of abnormally large elongated rutile grains would involve the crystal rotation during the coarsening process as reported elsewhere [40]. The abnormally large rutile grains appear to grow in all directions with little or no evidence of a preferred orientation.

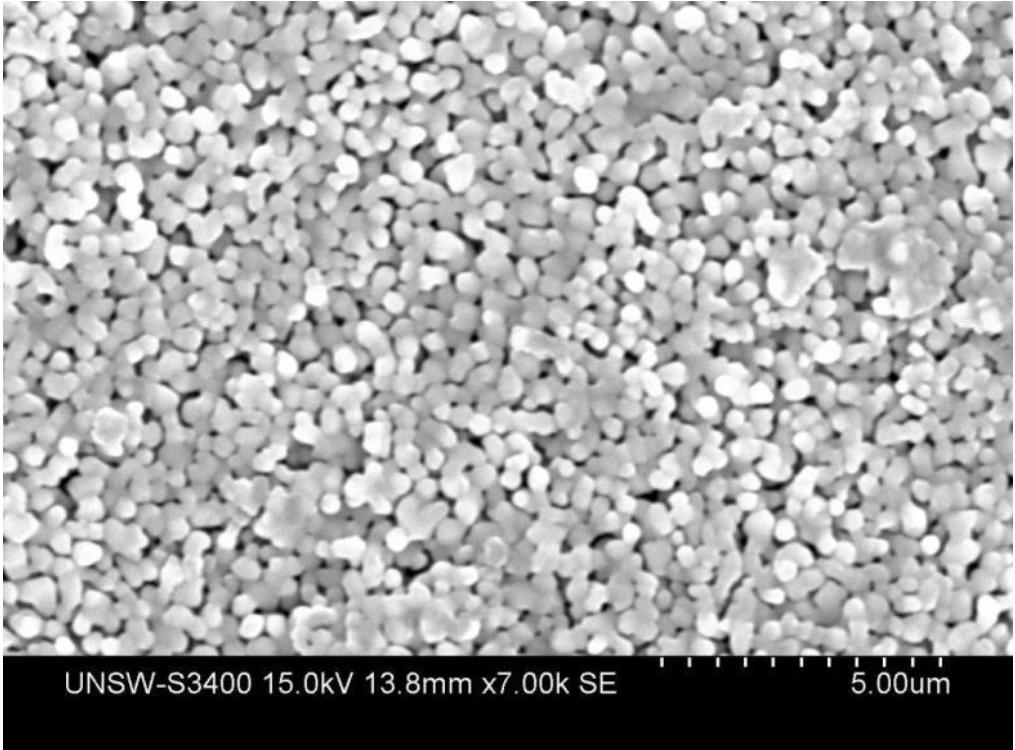

UNSW-S3400 15.0kV 13.8mm x7.00k SE          5.00um

a)

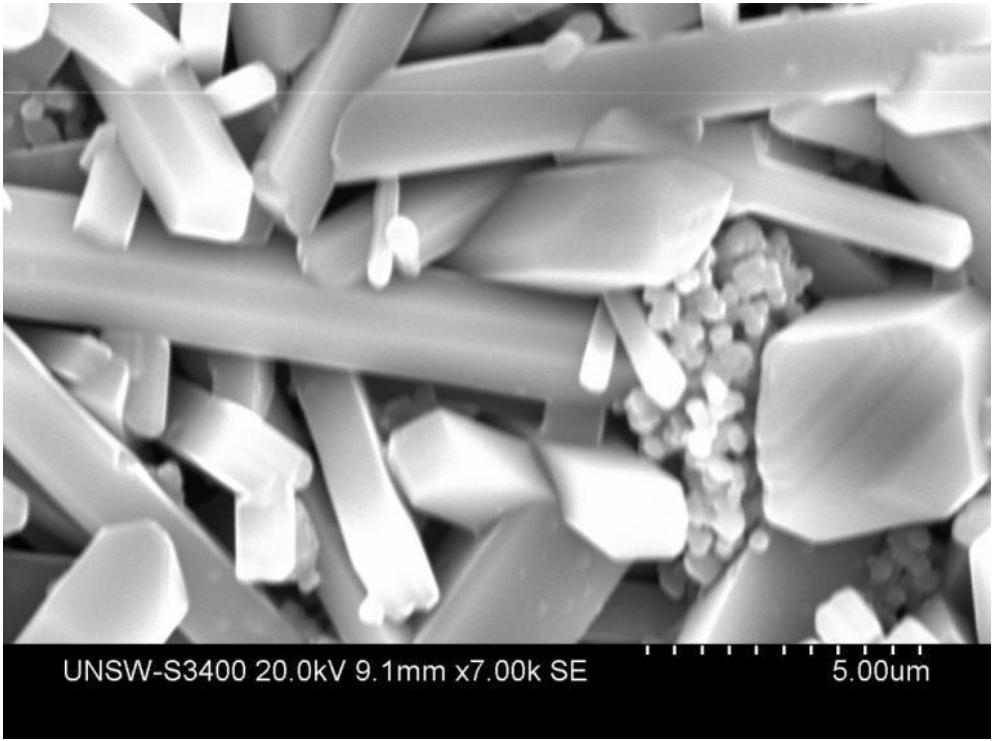

UNSW-S3400 20.0kV 9.1mm x7.00k SE          5.00um

b)

Fig. 2. SEM images of sintered $TiO_2$ fully transformed to rutile subsequent to firing at 1025°C for 12h (a) typical undoped sample exhibiting normal grain growth (b) Zircon doped sample (15wt%) exhibiting AGG.

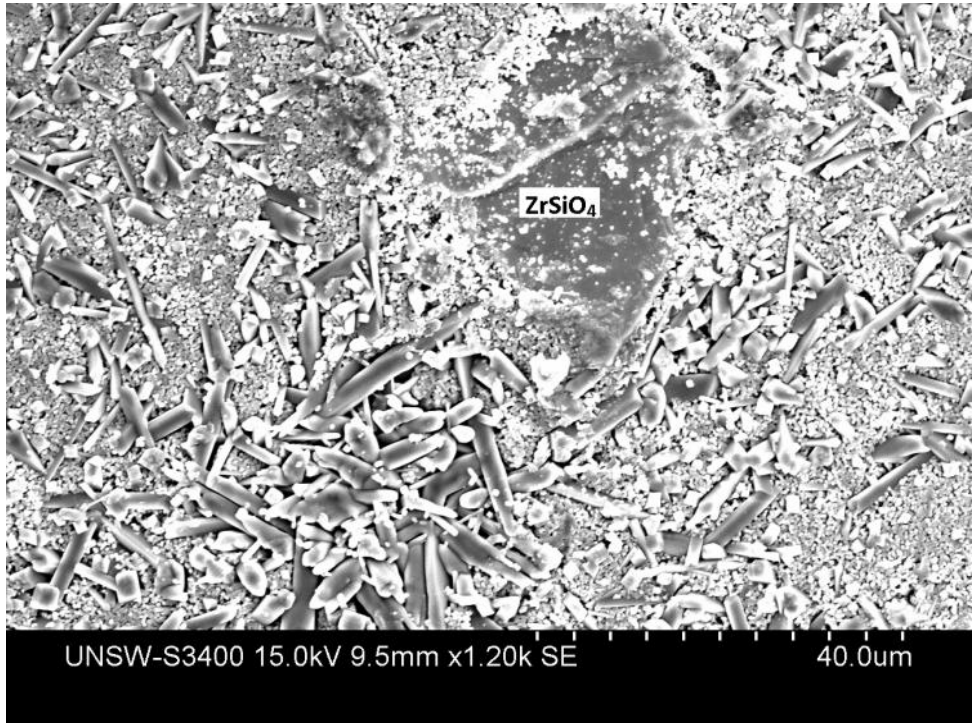

UNSW-S3400 15.0kV 9.5mm x1.20k SE          40.0um

**Fig. 3. Abnormal grain growth of rutile in the vicinity of a zircon particle (identified by EDS) in a sample doped with 10 wt.% ZrSiO$_4$ fired for 4h at 1025°C.**

Abnormally large needle-like grains were visible in optical micrographs. Raman spectra gathered from the regions of abnormal grain growth showed that these grains are of the rutile phase, while anatase is the dominant phase in the regions devoid of these grains. **Fig. 4** Shows an optical micrograph showing needles of rutile growing in the region surrounding a zircon particle the labels A, R and Z mark respectively anatase, rutile and zircon as identified by laser Raman microspectroscopy. It should be noted that Raman spectra are inappropriate for quantitative analysis and the anatase phase tends to exhibit stronger Raman shifts relative to rutile, as reported elsewhere [41, 42].

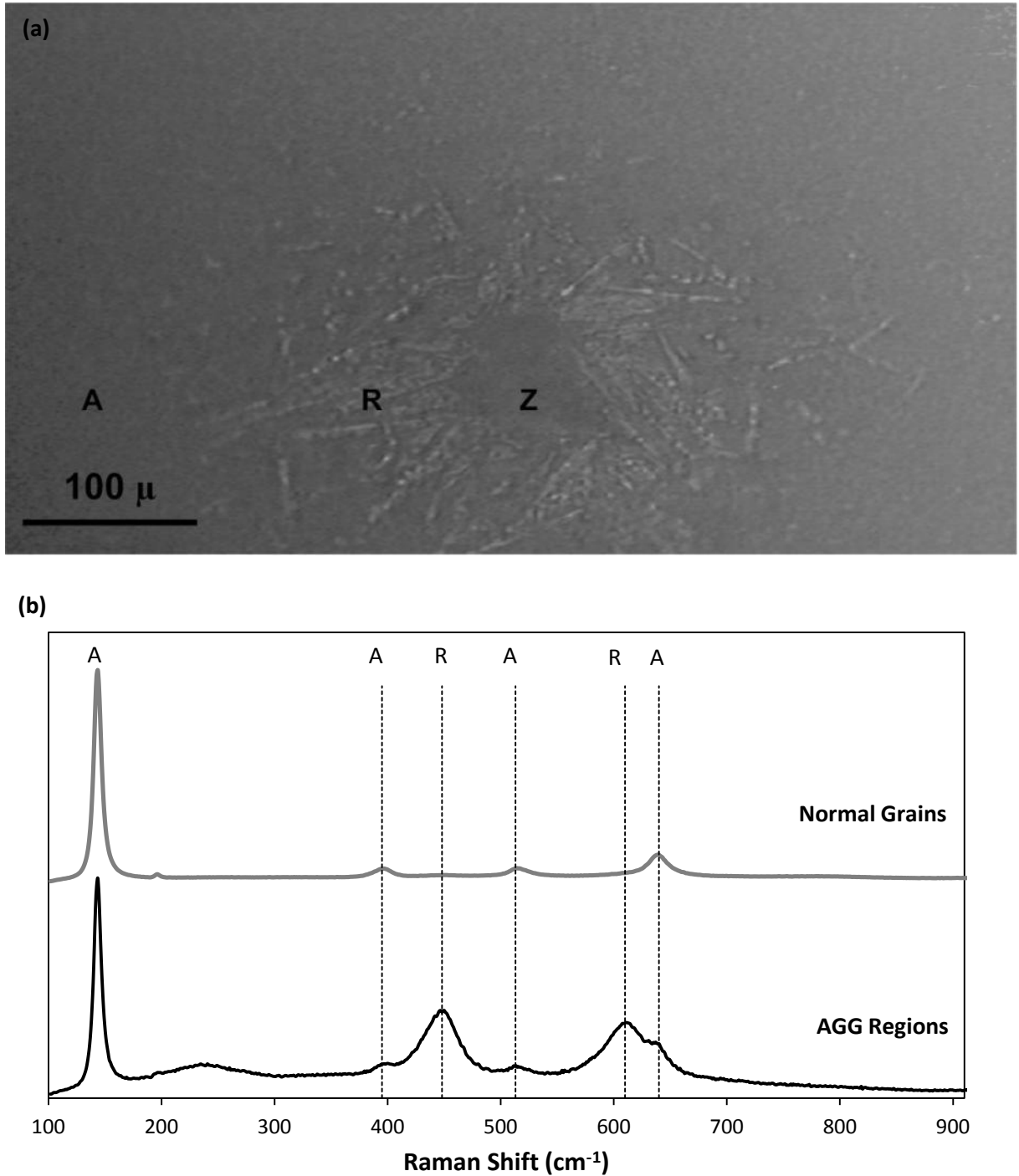

Fig. 4. (a) Optical microscope image of a zircon particle (Z) surrounded by AGG rutile in a 10 wt% ZrSiO$_4$ doped sample subsequent to firing at 1025°C for 4h (b) Raman patterns from regions of normal grain growth (A)  and AGG (R) in Fig. 4a.

 Table 1 outlines the impurity oxides which were present in the raw materials as determined by X-Ray fluorescence (XRF).

**Table 1.  Impurities in raw materials as determined by XRF**

| Raw Material | Impurity | Level (at%) |
|---|---|---|
| | $P_2O_5$ | 0.32 |
| | $Na_2O$ | 0.22 |
| Anatase | $K_2O$ | 0.21 |
| | MgO | 0.11 |
| | $SiO_2$ | 0.11 |
| | $HfO_2$ | 1.19 |
| | $Al_2O_3$ | 0.62 |
| Zircon Flour | $TiO_2$ | 0.30 |
| | $Fe_2O_3$ | 0.18 |
| | MgO | 0.08 |

In contrast to doping with $ZrSiO_4$, doping with $ZrO_2$ and $SiO_2$ does not lead to abnormal grain growth. This is demonstrated in **Fig. 5,** which depicts the morphology of $TiO_2$ in the region of a $ZrO_2$ particle, this is consistent with other studies which showed normal grain growth in $ZrO_2 - TiO_2$ and $SiO_2 - TiO_2$ systems [15, 37]. The presence of $ZrO_2$ particle was confirmed using EDS. Additions of $SiO_2$ powders and quartz sand to powder compacts also did not produce AGG.

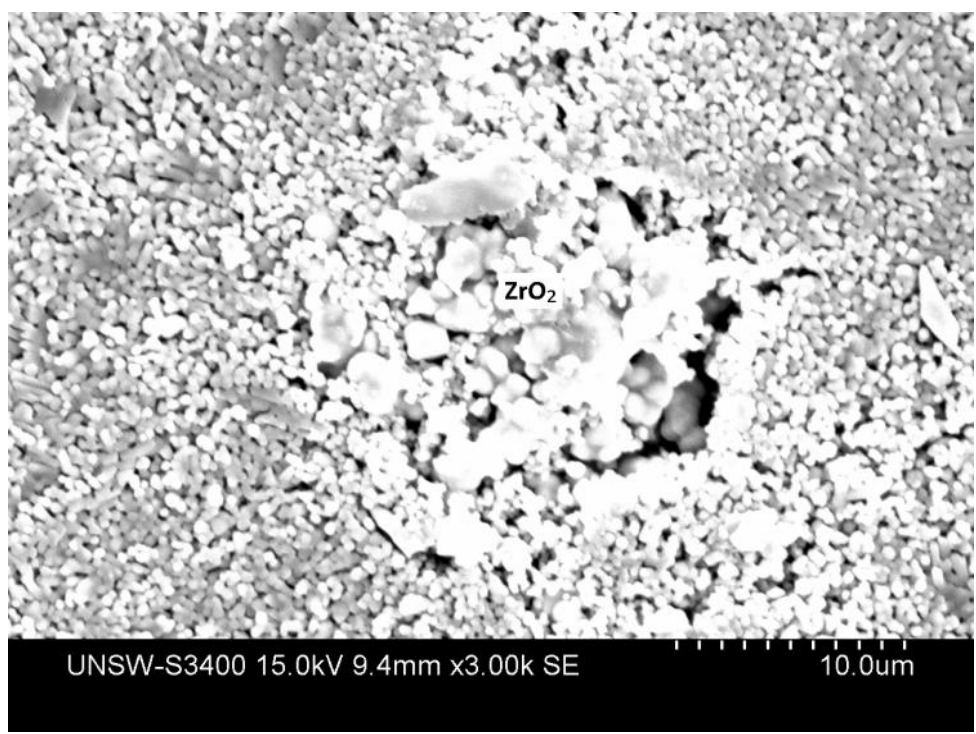

**Fig. 5. Morphology of $TiO_2$ surrounding an agglomerated particle of monoclinic $ZrO_2$ (identified by EDS) subsequent to firing at 1025°C.**

Further ruling out the influence dopant effects on the growth habit, the introduction of Si and Zr dopants using a wet impregnation method did not bring about abnormal grain growth as shown in the micrograph in **Fig. 6.**

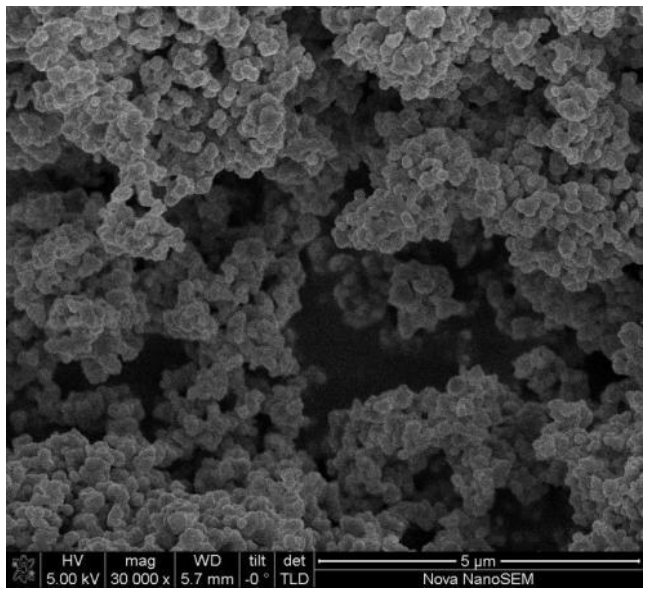

**Fig. 6. The microstructure of TiO$_2$ powder doped with 5 at% Si and 5 at% Zr using a wet impregnation method and subsequently fired at 1025°C.**

### 3.2.    Thin films

The use of Raman spectroscopy and glancing angle XRD confirmed thin films deposited by spin-coating on quartz substrates were fully transformed to the rutile phase after firing at 900°C. Undoped films did not exhibit AGG behaviour, as shown in **Fig. 7a.** As with other materials, here too the presence of zircon particles induced abnormal grain growth in the TiO$_2$ film. As shown in **Fig. 7b,** needle like grains of rutile were observed emanating from regions at or close to interfaces with particles of zircon. Zircon particles appear bright in SEM micrographs owing to the higher atomic number of Zr than Ti. Owing to the finer crystallite size of the precursor anatase, the acicular rutile grains in thin films were smaller than those observed in powders and varied in size from 0.1-3 μm in width and up to 20μm in length. The AGG of rutile appears to occur preferably in certain orientations. This is evident from the frequent occurrence of grains growing in parallel to each other and at 60° to each other.

It should be noted that abnormal grain growth was not observed in the vicinity of all zircon particles which is likely the result of varying levels of contact between the TiO$_2$ film and the zircon particulate inclusions.

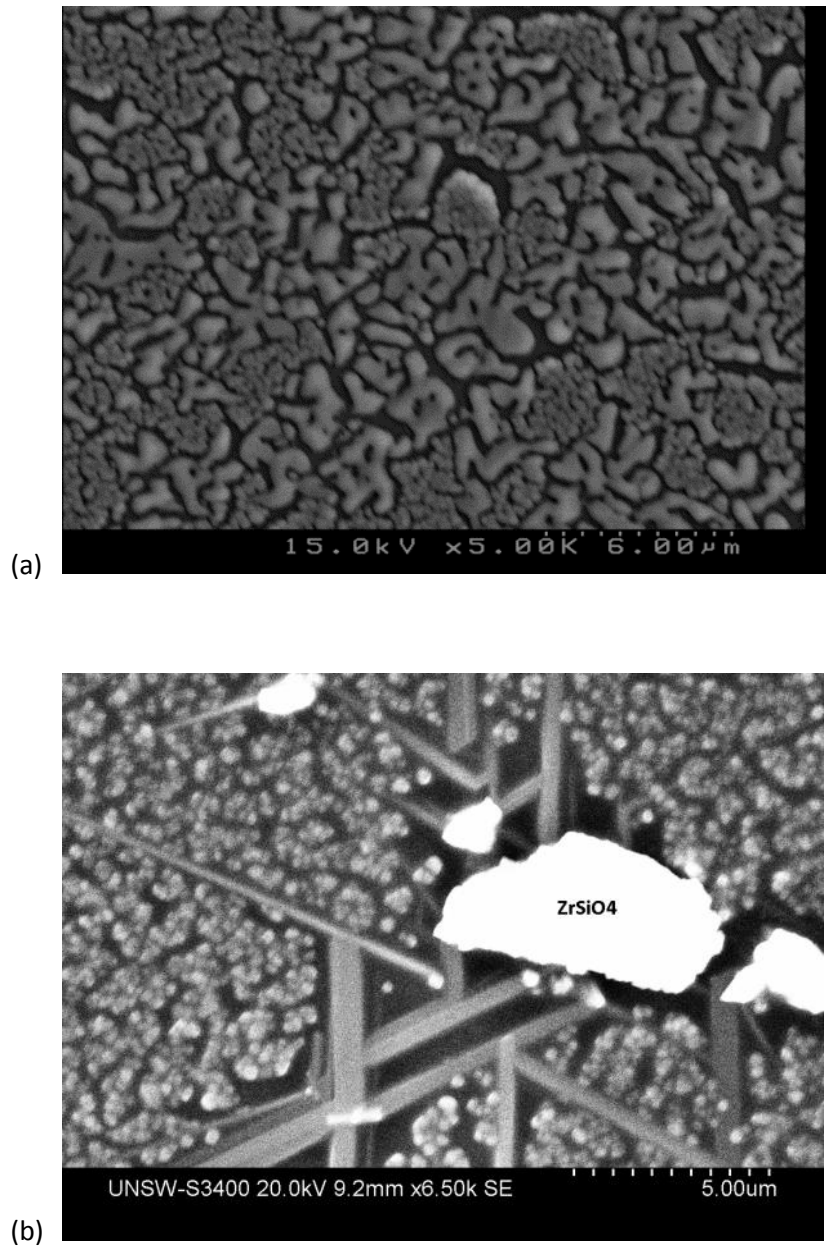

**Fig. 7. The morphologies of grains in thin films of rutile-TiO$_2$, deposited on quartz  substrates, after firing at 900°C for 12h. (a) undoped and (b) doped with ZrSiO$_4$ particles.**

### 3.3.  Coated Sands

Titanium dioxide coatings were fabricated on grains of sand as part of an effort to produce high surface area supported photocatalysts for water purification applications. It was observed that when zircon sand was used as a catalyst support material, coatings which had been fired at sufficiently high temperatures to form the rutile phase of TiO$_2$ exhibited randomly oriented elongated grains in a prismatic growth habit on certain facets of sand grains. This can be seen in **Fig. 8b.** In contrast, coatings of TiO$_2$ supported on quartz sand did not exhibit abnormal grain growth rather exhibited finer rounded rutile crystallites in the region of 50-500nm which varied in size with the thickness of the coating as shown in **Fig. 8a**. It should be noted that the morphology of coatings on zircon sand varied across different regions, with acicular rutile grains such as those visible in Fig. 8b observed only on certain facets of some grains. This suggests the crystallographic orientation of zircon plays an important role in giving rise to AGG $TiO_2$.

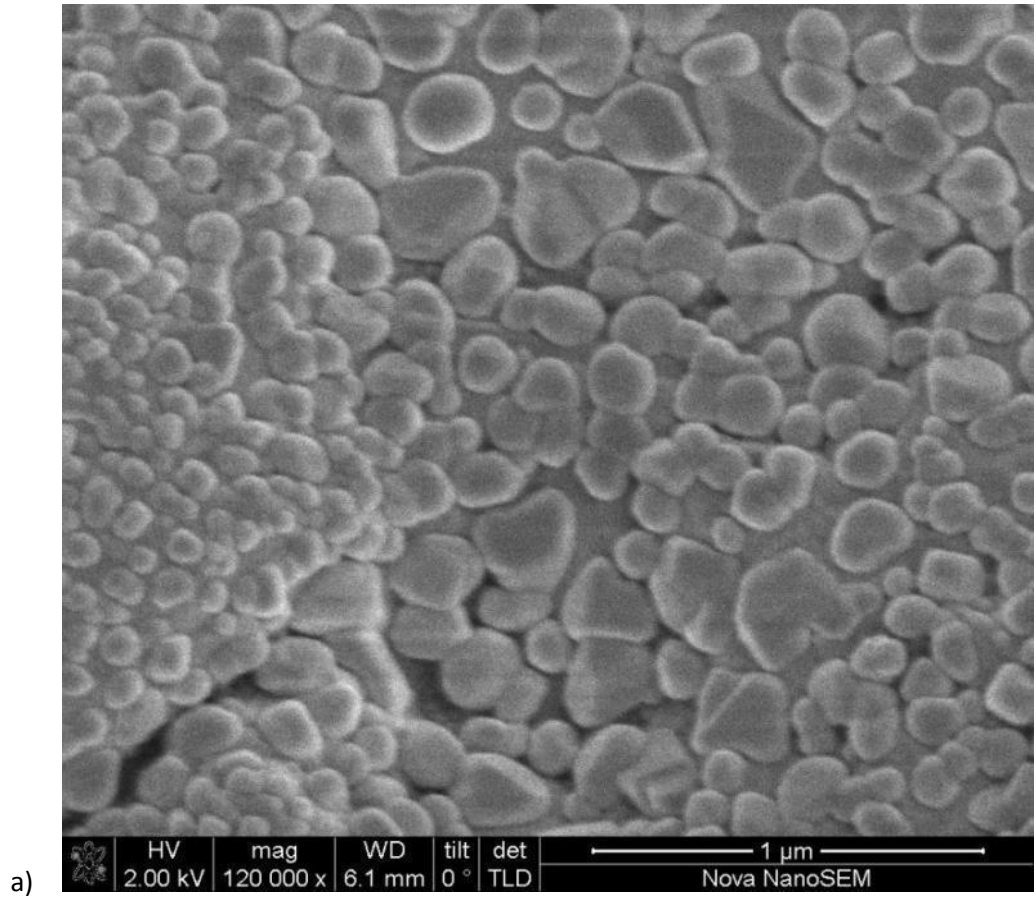

a)

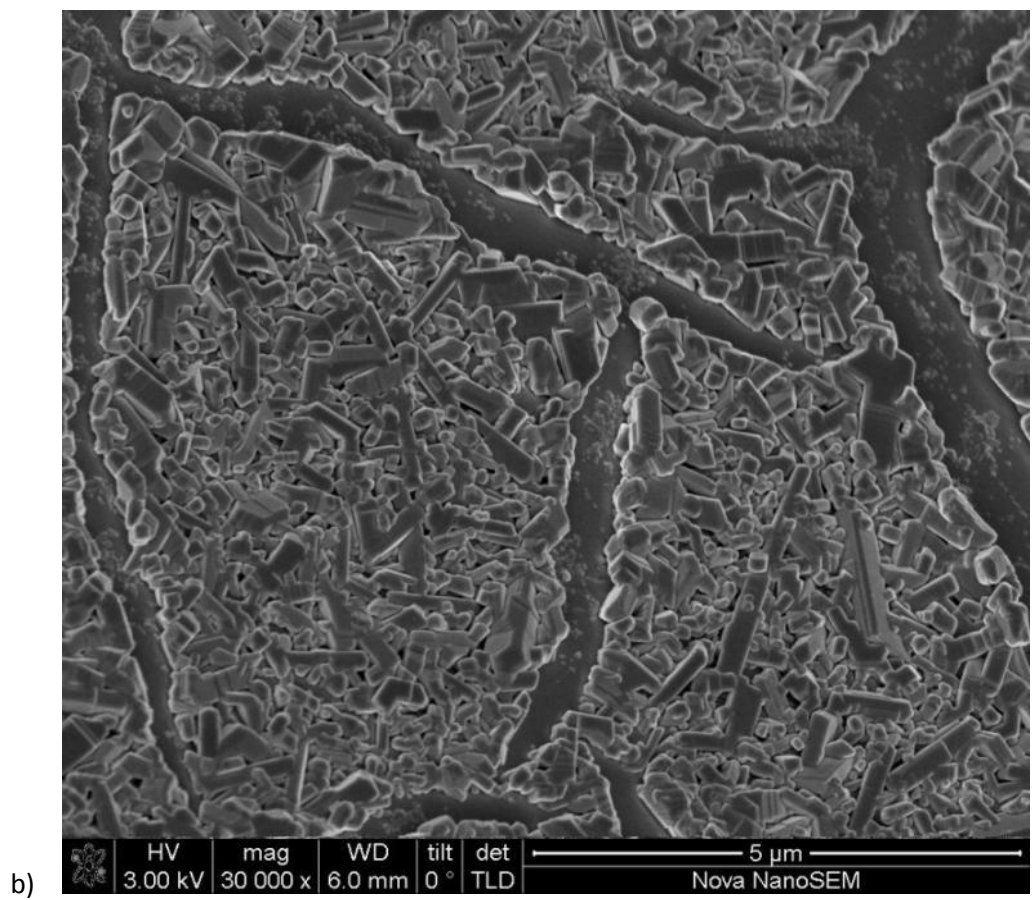

b)

**Fig. 8. Grain morphologies of sand-supported rutile TiO$_2$ coatings after firing at 1000°C for 4h**

**(a) on quartz sand and (b) on zircon sand.**

*3.4.  Growth habit of TiO$_2$ on crystal planes of zircon*

A single crystal of zircon was used to examine the growth habit of TiO$_2$ on different crystallographic planes of ZrSiO$_4$. XRD of small crushed and ground sample piece from the single crystal used confirmed the specimen to be of a zircon structure, although colour inducing contaminants were present.  After cutting the single crystal into several pieces and polishing the revealed faces, single crystal XRD was employed to determine the planes present in each section. These are outlined in table 2.

**Table 2. Rutile TiO$_2$ coatings on facets of a sectioned ZrSiO$_4$ single crystal**

| Piece | Plane(s) identified | Growth habit |
|-------|---------------------|--------------|
| A | (111) (010) | AGG |
| B | (122) | Normal |
| C | (403) | AGG/Normal |
| D | (011) | AGG |

| E | (001) | AGG |
|---|---|---|
| F | (012) (011) | Mostly normal |
| G | (011) | Mostly normal |
| H | (112) | AGG |
| I | (001) | AGG |
| J | (011)  (101) | AGG |
| Artificial Gemstone | (010) | AGG |

On single crystal $ZrSiO_4$ surfaces exhibiting AGG of $TiO_2$, large prismatic grains were frequently observed growing in preferred orientations, although smaller needle like grains were observed growing in random orientations. Abnormal grain growth of $TiO_2$ on the majority of surfaces suggests a particular orientation of $ZrSiO_4$ might not be a prerequisite for this type of grain growth in $TiO_2$.

### 3.5.    Crystallographic orientation relationship between $TiO_2$ and $ZrSiO_4$

Rutile is known to exhibit preferential growth along the c-axis owing to the lower surface energies of planes exposed in this growth habit. This has been demonstrated through the use of Wulff constructions [43]. The $ZrSiO_4$ induced growth of abnormal rutile-$TiO_2$ grains elongated along the c-axis, as seen in Fig. 2b, may be the result of an orientation relationship between the product rutile phase and zircon particles that reduces activation energy for rutile nucleation on zircon particles. This relationship may be predicted according to various models, such as edge-edge matching model, based on the principle of lattice structure and parameter matches of these two phases [44]. The best possible orientation relationship between rutile and zircon particle is predicted as following, based on the matching principle, and the lattice structures and parameters of rutile and zircon (Table 3).

$$[001]_{ZrSiO_4} //[001]_{r-TiO_2} , (020)_{ZrSiO_4} //(110)_{r-TiO_2}$$

The interplanar spacings (d-value) for the two close-packed matching planes of these two phases are 0.3302 nm for $(020)_{ZrSiO_4}$ and 0.3248 nm for $(110)_{r-TiO_2}$, and the interatomic spacing along the close-packed atomic rows are 0.299 nm for $[001]_{ZrSiO_4}$ and 0.296 nm for $[001]_{r-TiO_2}$. The interplanar spacing (d-value) mismatch between the matching planes, and the interatomic spacing misfit along the matching directions, of these two phases, is 1.6% and 1.0% respectively; these are well below the critical values 6% for interplanar spacing mismatch and 10% for the interatomic spacing misfit of the two matching phases in the edge-edge matching model of Kelly and Zhang [44]. **Fig. 9.** clearly shows a near perfect match of Ti, Zr and Si atoms in the two matching $(020)_{ZrSiO_4}$ and $(110)_{r-TiO_2}$ planes along the matching directions of $[001]_{ZrSiO_4}$ and $[001]_{r-TiO_2}$ in this orientation relationship.

**Table 3: Crystallographic properties of zircon rutile and anatase**

| Phase | Stability | Crystal Structure | Space Group | Lattice Parameters Å | Atoms per unit cell |
|---|---|---|---|---|---|
| Zircon | Equilibrium Phase up to 1676 °C [45, 46] (decomposition) | Tetragonal | I4$_1$/amd | a=6.6; c=5.98 | 4 |
| Rutile | Equilibrium Phase up to 1870 °C [47-49] (congruent melting) | Tetragonal | P 42/mnm | a=4.6; c=2.96 | 2 |
| Anatase | Metastable Phase Transformation to rutile ~650°C [42, 50] | Tetragonal | I4$_1$/amd | a=3.8; c=9.51 | 4 |

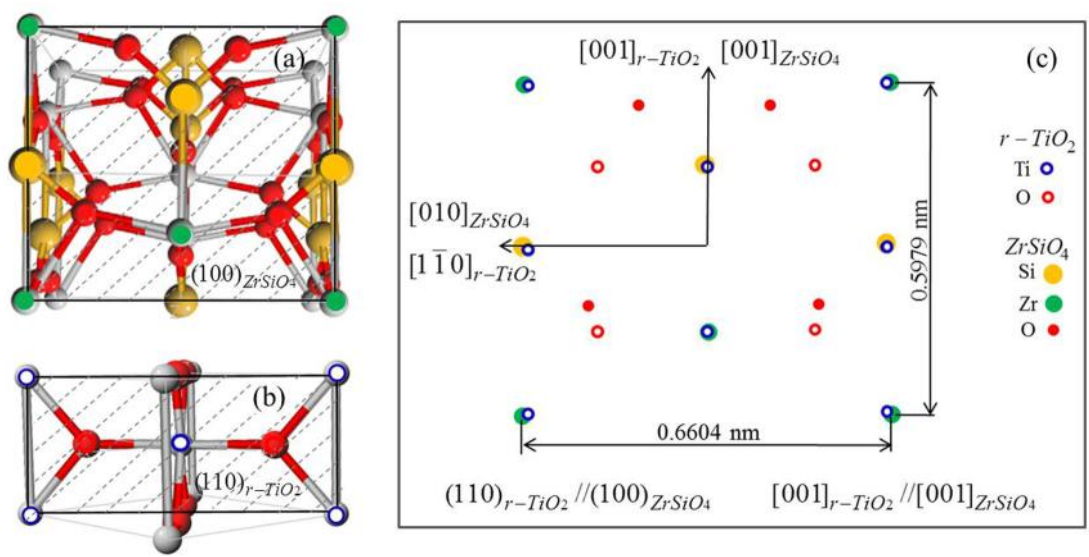

**Fig. 9:  Diagrams showing the lattice structures of (a) zircon and (b) rutile, and (c) the matching of Zr, Si and Ti atoms in the close-packed matching planes of** $(020)_{ZrSiO_4}$ **and** $(110)_{r-TiO_2}$ **along the close-packed matching directions of** $[001]_{ZrSiO_4}$ **and** $[001]_{r-TiO_2}$ **, according to the predicted orientation relationship between zircon and rutile.**

## 4.  Discussion

While the formation of elongated rutile crystallites has been reported previously, this has typically involved the formation of nano-scale rutile crystallites from precursor solutions using hydrothermal synthesis routes [51-53]. Or through the intermediate formation of sodium titanate compounds [54, 55]. The abnormal grain growth of rutile TiO$_2$ from a polycrystalline anatase precursor as shown in this work has not been previously reported.

Abnormal grain growth resemblant of that observed in this work has been reported in various ceramic systems including $Al_2O_3$ [3, 6], 0.1 mol% $TiO_2$ excess $BaTiO_3$ [7, 8], $Si_3N_4$ [56], and $(Na,K)NbO_3$ piezoelectric ceramics [57]. This abnormal grain growth involves the rapid anisotropic growth of certain grains in comparison with surrounding material. The conditions which have been reported to give rise to such growth include high anisotropy in interfacial energy, high chemical inequilibrium and/or an non-uniform distribution of second phase particles [10]. Within single phase systems faceted grain boundaries are frequently reported as a prerequisite for the onset of AGG. We can consider these conditions in reference to the systems studied in the present work.

### 4.1.    Epitaxial nucleation

The elemental diffusion of Si or Zr has been shown not to cause AGG in $TiO_2$. Thus the initiation of AGG at zircon surfaces suggests a crystallographic nucleation effect. Furthermore, in coatings fabricated on grains of zircon sand, large prismatic rutile grains were only observed on particular facets of sand, suggesting a particular orientation is necessary for this phenomenon to occur.

An orientation relationship between zircon and rutile was found to involve a high level of matching between values of interatomic and interplanar spacing of the two structures.  This relationship satisfies the conditions for epitaxial nucleation and growth as set out by Kelly and Zhang [44].

The existence of an orientation relationship between zircon and rutile is supported by the angles between AGG rutile crystallites seen in zircon doped $TiO_2$ films on quartz substrates, shown in Fig. 7b, which are frequently close to 0, 60 or 120°. Such angles occur also in $TiO_2$ coatings on zircon sand, as shown Fig 8b, although to a lesser extent. Such growth angles are common in cubic systems rather than the tetragonal structures present in the investigated systems, it is however possible that these preferred orientations result from substrate effects (a possible orientation relationship with the single crystal quartz substrate).

In contrast to the evidence supporting crystallographic nucleation phenomena, the largely random orientation of AGG grains in powder compacts does not consistently conform to what would be expected as the result of nucleation due to an orientation relationship between zircon particles and abnormally large prismatic grains of rutile. This suggests other factors may well be involved in inducing the observed growth.

AGG of rutile $TiO_2$ as a result of an epitaxial orientation relationship with zircon has not been conclusively determined and further work utilising Electron Backscatter Diffraction (EBSD) and TEM is likely to shed further light on the observed phenomena reported here.

### 4.2.    Distribution and size of second phase particles

 In a publication on the theory of normal and abnormal grain growth, Hillert [10] considered 2 and 3 dimensional systems containing pinning second phase particles, here it was suggested that AGG occurs when the average grain size in a material lies within a particular interval relative to the volume fraction and size of the second phase particles.

From analysis of densely packed hexagonal grains with uniformly distributed growth restricting particles of constant size, AGG was predicted to arise when the average grain size $\overline{R}$ is in the range:

$$\frac{4r}{9f} < \overline{R} < \frac{2r}{3f} \qquad (1)$$

Here $r$ and $f$ are respectively the radius and volume fraction of the second phase particles.

This analysis is unlikely to be applicable to the $TiO_2$ systems investigated in this work, which are significantly divergent from the dense metallic grain structures of Hillert's analysis. Furthermore, owing to the large sizes of the zircon secondary particles in comparison with the $TiO_2$ grains, the condition of Eq. 1. Is not met and also the occurrence of AGG in $TiO_2$ coatings on the surfaces grains of zircon sand single crystals cannot be explained by this mechanism. Additionally, if a simple grain-edge pinning mechanism was sufficient to cause AGG in $TiO_2$, this phenomenon would occur with a wide range of other stable solid state dopants, contrary to what has been observed.

### 4.3. Chemical inequilibrium and pinning forces

A high degree of inequilibrium is reported to facilitate AGG in single phase systems. It has been reported that in systems where grain growth is pinned, abnormal grain growth may occur if the pinning force is rapidly lessened [2]. The applicability of these observations to a ceramic system such as that studied in the present work is unclear. Anatase is a metastable phase with inequilibrium increasing as temperature is increased [58-60]. It should be noted however that abnormal grain growth is not observed other systems where the anatase to rutile phase transformation temperature is increased through transformation inhibiting oxide dopants [37, 42, 61, 62]. It is nonetheless possible that an increased level of chemical inequilibrium brought about through the restriction of the anatase to rutile phase transformation plays a role in the AGG reported in the present work.

### 4.4. Surface energy anisotropy

The anisotropy of surface energy in $TiO_2$ is well documented. It is reported that in rutile (110) planes consistently exhibit lower surface energy relative to other rutile planes [43]. This explains why rutile frequently forms in a prismatic or acicular growth habit extended along the c-axis [63]. The divergence of surface energies of the different planes varies with the type of surface terminating groups and it has been reported that under strongly basic conditions of surface acidity (oxygenated surfaces) the elongation of rutile grains is most pronounced as a result of larger disparity in surface energies under such conditions [43, 64].

Surface energy anisotropy in rutile has not been reported to give rise to abnormal grain growth in polycrystalline $TiO_2$ such as that observed in this work. Furthermore, in the absence of zircon dopants, isotropic grains of anatase, in the various forms studied in the present work, were observed to transform into larger isotropic grains of rutile rather than elongated prismatic grains.

One could consider the hypothesis that the presence of zircon particles alters the surface chemistry of surrounding $TiO_2$ grains thus increasing the driving force for the formation of prismatic grains. This theory is unlikely as $ZrSiO_4$ surfaces do not exhibit strong surface basicity.

### 4.5.    Fluxing impurities

The presence of a liquid phase at grain boundaries can be a contributing factor to the anisotropy in surface energy and thus give rise to the rapid growth along a particular crystallographic direction, giving rise to AGG [1, 3]. From consideration of phase equilibrium diagrams, the formation of liquid in the $TiO_2$-$ZrSiO_4$ system is not expected to occur at the temperatures employed in the present work [65, 66]. However, the presence of fluxing impurities should be considered, particularly as the zircon flour used in this work exhibits ~0.3 at% Fe impurity according to the supplier's specifications. Studies which have investigated the $FeO$-$ZrO_2$-$SiO_2$ system have shown liquid formation at temperatures lower than the $ZrSiO_4$ decomposition temperature at compositions with  > 5%Fe levels [67, 68], however solidus temperatures reported are still significantly higher than those used in the present work. Furthermore, AGG was observed in $TiO_2$ coatings on the (010) surface of an artificial $ZrSiO_4$ gemstone, which did not exhibit any coloration, suggesting no significant levels of transition metal impurities. In spite of this, based on the appearance of the propagation of AGG in the vicinity of zircon particles, grain boundary liquid formation appears a likely mechanistic candidate. The presence of low levels of alkali contaminants may be a contributing factor to this grain boundary liquid formation.

### 4.6.    Faceted grain boundaries

The morphology of the precursor anatase material is crucial in determining whether rutile exhibiting abnormal grain growth will result from heat treatment. In reference to systems other than that studied in the present work, faceted grain boundaries have been reported to be a prerequisite for the occurrence of AGG [8, 69]. Thus rounded grains or grains exhibiting atomically rough or defaceted grain boundaries are not expected to exhibit AGG.

The random orientation of abnormally large prismatic rutile in the vicinity of zircon particles is inconsistent with the theory that epitaxial nucleation is the driving force for this grain growth as nucleation effects would involve a particular orientation relationship and one would expect to see preferred orientations of AGG rutile relative to the zircon particles. An alternative explanation may be the rutile $TiO_2$-$ZrSiO_4$ interface acts as a faceted grain boundary, leading to the localised abnormal growth of more rutile grains which themselves are faceted and in turn cause further AGG.

### 4.7.    Combined effects

Whilst determining a distinct mechanism by which $ZrSiO_4$ imparts AGG in rutile $TiO_2$ is somewhat controversial, the observed phenomena are potentially the result of the cumulative effects of different parameters which may include:

- Grain size of precursor anatase $TiO_2$
- Crystallographic orientation relationship between $TiO_2$ and certain planes of zircon $ZrSiO_4$
- The presence of fluxing impurities
- Pinning particles and an increased chemical inequilibrium due to the restriction of the anatase to rutile phase transformation

- Surface acidity

These factors may be cumulatively involved in inducing AGG in rutile $TiO_2$. Abnormally large faceted grains may then propagate throughout the bulk of the material through nucleation of prismatic grains at faceted boundaries.

## Conclusions

The rutile phase of $TiO_2$ has been shown to exhibit abnormal grain growth when formed by the thermal treatment of anatase in the presence of zircon, $ZrSiO_4$. Along with other factors, a distinct relationship between the crystallographies of the two materials, involving a high degree of atomic position matching, is likely to play an important role in inducing this type of grain growth. It is likely that the abnormal grain growth observed in this work can be achieved in $TiO_2$ with the use of dopants other than zircon. Further studies are required to elucidate the mechanisms which can give rise to AGG in $TiO_2$ and examine the properties of materials exhibiting this type of grain growth in order to apply $TiO_2$ materials exhibiting the observed phenomena.